# LEGITIMIZATION OF DATA QUALITY PRACTICES IN HEALTH MANAGEMENT INFORMATION SYSTEMS USING DHIS2. CASE OF MALAWI.

Martin Bright Msendema, University of Malawi, mmsendema@poly.ac.mw

Wallace Chigona, University of Cape Town, wallace.chigona@uct.ac.za

Benjamin Kumwenda, University of Malawi, benjkum@gmail.com

Jens Kaasbøll, University of Oslo, jensj@ifi.uio.no

Chipo Kanjo, University of Malawi, chipo.kanjo@gmail.com

**Abstract:** Medical doctors consider data quality management a secondary priority when delivering health care. Medical practitioners find data quality management practices intrusive to their operations. Using Health Management Information System (HMIS) that uses DHIS2 platform, our qualitative case study establishes that isomorphism leads to legitimization of data quality management practices among health practitioners and subsequently data quality. This case study employed the methods of observation, semi structured interviews and review of artefacts to explore how through isomorphic processes data quality management practices are legitimized among the stakeholders. Data was collected from Ministry of Health's (Malawi) HMIS Technical Working Group members in Lilongwe and from medical practitioners and data clerks in Thyolo district. From the findings we noted that mimetic isomorphism led to moral and pragmatic legitimacy while and normative isomorphism led to cognitive legitimacy within the HMIS structure and helped to attain correctness and timeliness of the data and reports respectively. Through this understanding we firstly contribute to literature on organizational issues in IS research. Secondly, we contribute to practice as we motivate health service managers to capitalize on isomorphic forces to help legitimization of data quality management practices among health practitioners.

## 1. INTRODUCTION

Within the neo-institutional thinking we learn of the role of legitimacy and isomorphic forces in trying to understand success and changes in organizations which are attained by conforming to constraints within their operating environments (DiMaggio & Powell, 1983). The term legitimacy means a generalized perception or assumption that actions of an entity are desirable, proper, or appropriate within some socially construed systems or norms, values, beliefs, and definitions (Lebbadi, 2015). We share the understanding of isomorphism as a constraining process that forces one unit in a population, like an organization, to resemble other units that face same set of environmental conditions (Currie, 2012). The interplay between isomorphism and legitimacy is experienced in such a way that in the process of isomorphism institutions obtain legitimacy (Díez-Martín, Díez-de-Castro, & Vázquez-Sánchez, 2018). The work of Freitas & Guimarães (2007) demonstrates how various isomorphic forces lead to various types of legitimacy in organizations. In their study they found that isomorphism led to cognitive legitimacy in the organization they were studying.

We hence envisage that if Ministry of Health in Malawi legitimize data quality management practices (data collection, processing, reporting, monitoring and use) through isomorphism across the players in Health Management Information System (HMIS), a homogeneous behaviour which is conforming to data quality management practices can be achieved among the stakeholders leading to improvement in data quality for all the players.

The DHIS2 has been designated as a platform to support data collection, analysis, storage and use with the hope of attaining data quality in the Malawi HMIS (Ministry of Health -Malawi, 2018). Data quality is a common discourse in routine health information systems research. Poor data quality is a serious concern in many disciplines including health, where it often leads to the loss of innocent lives (Mettler, Rohner, & Baacke, 2008). Within this debate, the definition of data quality has been one of the interesting topics. There have been different schools of thoughts (Chen, Hailey, Wang, & Yu, 2014 ). Nonetheless, the notion that data quality relates to how the data satisfy the expectations of the user or how it fits with the aims of its use has been hugely accepted among scholars (Harpe & Roode, 2008; Chen, Hailey, Wang, & Yu, 2014 ). This study hence adopts that definition. Although implementation of DHIS2 has contributed to tangible success, recent studies have reported that political, social and cultural issues are among the challenges affecting the successful implementation of the DHIS2 across the developing countries (Dehnavieh, et al., 2018). For this reason, we found that DHIS2 is a compelling example to use when attempting to explain how understanding of the interplay between legitimacy and isomorphism may contribute to actions to improve data quality in HMIS.

Diversity of stakeholders in health information management in Malawi creates a complex HMIS institution that is difficult to manage. In most developing countries, the governments are supported by health partners who have specific objectives and modus operandi creating a battleground (Currie, 2012). In Malawi, within the data processing chain the government (Ministry of Health) is the data owner with its staff (medical, management including ICT support and statisticians). Other actors include health partners who provide technical staff (ICT specialists) and medical staff (programme coordinators). Basing on a definition of institutions as the humanly devised constraints both formal (rules or laws) and informal (norms of behaviour and codes of conducts) that structure human interaction and their enforcement mechanisms (Wang & Ching, 2013), then we can perceive HMIS as an institution; HMIS has its own practices, rules and norms (Kimaro, 2006). However, the challenge with Malawi's HMIS is that it has multiple players. Such an environment expects all the stakeholders to conform to these constraints regardless that they may each have individual characteristics and expectations.

One of the problems that Malawi is experiencing from having multiple stakeholders in HMIS is that other stakeholders do manage to get the expected quality data while others still struggle to attain high levels of data quality. Among those reported with low data quality problem, studies have shown that timeliness and correctness continue to be their concern (O'Hagan, et al., 2017; Statistics Norway, 2017). Such a problem may lead to other health interventions being successful while others failing as their plans and decisions may not be based on reliable data. In the long run this deprives Malawi of the opportunity to meet the Sustainable Development Goal 3 whose target is 'Good Health and Well Being for all' (United Nations, 2017). This makes us contemplate on the legitimization of the data quality management practices among the various stakeholders as one of concepts that could help explain the phenomenon. Our curiosity is derived from the argument that legitimacy is key to successful implementation of difficult polices in organization (Wang & Ching, 2013, p. 523). Researchers in Malawi's HMIS have pointed at availability of resources and incentives as the reason behind differences in data quality between programmes under health partners and those not under direct interest of the health partners. Much as we appreciate that resources and incentives may influence adherence to data quality management practices, we fear that different norms (formal and informal) among the stakeholders contribute to different ways of legitimization among the players. Hence this study attempts to explain how data quality management practices attain legitimacy in HMIS. We posit that the findings can help practitioners to know how to manage the legitimacies towards high data quality across the entire HMIS. Reflecting on the argument that there is dearth of research focusing on how social issues affect success of information systems like the HMIS (Avgerou , The significance of context in information systems and organizational change, 2001; Aqil, Lippeveld, & Hozumi, 2009), we believe the study findings contribute towards a social perspective of actions to improve data quality in HMIS (Avgerou , The significance of context in information systems and organizational change, 2001; Aqil, Lippeveld, & Hozumi, 2009).

From this background, we build on the two important concepts, legitimacy and isomorphism and attempt to answer the question:

*What role do medical professional formal and informal norms play in operationalization of data quality management practices in HMIS that are using DHIS2?*

## 2. STUDY SETTING

### 2.1. DEMOGRAPHIC PERSPECTIVE

The 2018 census report indicates that Malawi's population is near 20 million of which 84% live in rural areas (National Statistical Office, 2019). Within the Southern Africa Development Community (SADC) region Malawi is argued to have the lowest Gross Domestic Product (GDP) of $430. It has an economy is agro based economy f which agriculture contributes about 28% to the GDP (Southern African Development Community, 2020). With the majority of the population living in the rural areas, they rely on free health care service from the government (Makwero, 2018). HIV and Aids, Tuberculosis, and malaria remain a significant challenge to health serving delivery with HIV leading on major causes of death at around 25% of the deaths. Malawi is probably one of the most donor supported health system with donors funding 80% of its development expenditure which in 2019 was about USD135 million (UNICEF, 2017; Ministry of Finance, 2019).

### 2.2. MALAWI HEALTH SYSTEM

Malawi follows a primary health care approach which is a whole-of-society approach to health and wellbeing centered on the needs and preferences of individuals, families and communities (Makwero, 2018; World Health Organization, 2019). The health system has a four-level structure comprising of community level, primary level, secondary level and tertiary level. Beyond the service delivery levels are Directors of Health and Social Services (formerly District Health Officers) which are responsible for management of health services and facilities including providing direction at the district. The Ministry of Health Headquarters is the central government's agent for policy making, standards setting, quality assurance, strategic planning, resource mobilization, technical support, monitoring and evaluation and international representation (Ministry of Health, 2016).

### 2.3. Health Management Information Systems in Malawi

HMIS implementation structure for Malawi has five levels: Community, facility, district, zone and national. This reflects on the national administrative organization of the health system structure (recall Section 2.2). The need to include informal sources of data as advocated by researchers such as Kanjo, Kaasboll and Sahay (2012) has seen data collection starting from the community (village) level, which is the first level of the five in the health system. Health Surveillance Assistants (HSAs) entrusted with data collection at this level, collect data manually from all possible health indicators on services acquired in the local community. After the community level, there is the facility level (health center). Data clerks and focal person collect data at health centers and hospitals, see Figure 1. The focal persons (who are nurses or clinicians) and data clerks (statisticians) compile the data manually on paper forms. At district hospital final reports consists of data from both the hospital and from all the health centers in the district (Chikumba, 2017). After reports are compiled, they are captured into DHIS2 and at each district level, there is an HMIS officer who oversees data management issues.

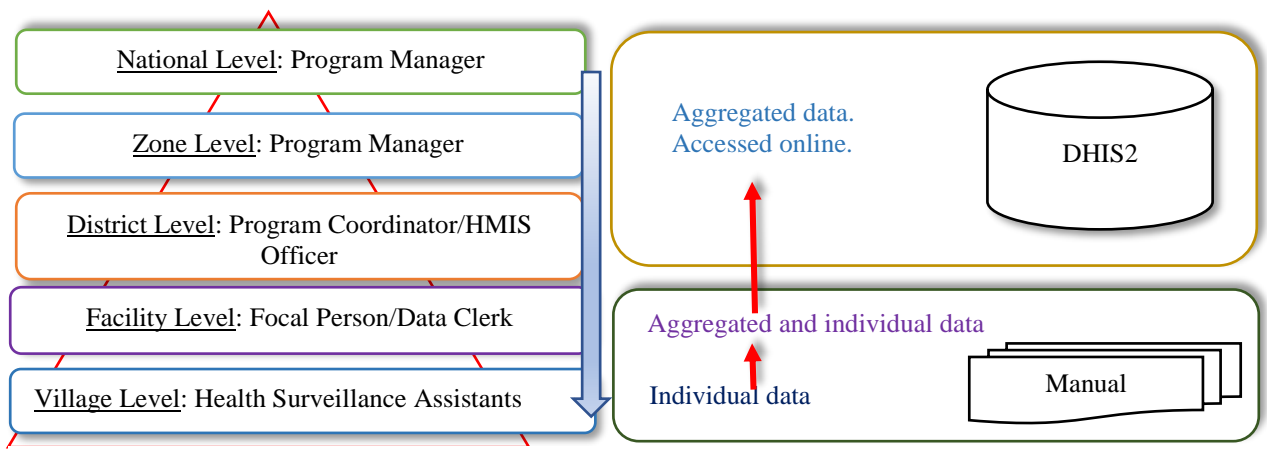

*Figure 1: HMIS implementation structure in Malawi-Source. Adapted (Chikumba, 2017; Wu, Kagoli, Kaasbøll, & Bjune, 2018)*

## 3. THEORETICAL FRAMEWORK

We start the theoretical framing of the study on the premise that "the idea of legitimacy is key to successful implementation of difficult policies and that any institution is legitimate when a large part of its target population recognizes and obeys the normative incentives set out by this institution" (Wang & Ching, 2013, p. 523). Within institutional theory it is argued that legitimacy must be granted by the governing institutions like the Ministry of Health and the Health Partners or local members of the environment like the HSAs. This can be achieved if the stakeholders realise instrumental value of organizational activities (pragmatic legitimacy), recognize the activity as morally good or there is ethicality of organizational activities (moral legitimacy) or acceptance of an activity as essential or permanent part of their behaviour (cognitive legitimacy) (Randrianasolo & Randrianasolo, 2017) (Bridwell-Mitchell & Mezias, 2012; Randrianasolo & Randrianasolo, 2017)

Now to ensure a large part recognizes the legitimacy, institution theory further provide mechanisms called isomorphic forces. Isomorphic forces / pressures are pressures or decisions by organizational actors that lead the organization to adopt structures, procedures, systems and terminology shared by other organizations of the same type (ECDPM, 2005). They are forces that lead to heterogeneous behaviour among stakeholders but all resulting into the legitimateness of data quality management practices. These forces are coercive (from rules, regulations), mimetic (occurring from uncertainties where behaviors are just imitations) and normative (from established patterns, policy, standards or training) (Currie, 2012).

Through these institutional theoretical concepts (legitimacy and isomorphism), we argue that isomorphic forces could play crucial role in helping us understand how the target population in HMIS, **see Table 1**, get to legitimize data quality management practices under the influence of isomorphism (Sampaio de Freitas & Guimarães, 2007).

### 3.1. STUDY LOCATION AND PARTICIPANTS

This is part of a longitudinal study which started in November 2020 and is expected to finish in May 2021. We first collected data in Lilongwe which is the central administrative district and later in Thyolo district. The study participants were drawn from four groups: (i) HMIS Technical Working Group (TWG), (ii) Hospital/facility Managers, (iii) Data Managers and (ii) Medical practitioners. The specific composition of the participants is presented in **Table 1**.

| Group | Participants |
|---|---|
| **HMIS Technical Working Group** | CMED(1) , DHO(1), HMIS OFFICE(1) |
| **Facility Managers** | Health Center In-charge (2), |
| **Medical practitioners** | Focal Persons (nurses, Clinicians -3), HSAs (10) |
| **Data Managers** | HMIS Officer, Data Clerk(2) |

*Table 1: Specific composition of the study participants*

### 3.2. STUDY APPROACH AND METHODS

This was a qualitative case study and used observation, semi structured interviews and study of artifacts as data collection tools. There was no specific sequence but multiple sources helped to triangulate and validate the results. Semi structured interviews involved members of the HMIS TWG at the Ministry of Health Headquarters, District Health Administrators including HMIS Officer, Data Clerks and Health Facility In-charge. This is a strategic level group and the focus was supervision and feedback, training and support and relationship in the various roles. Despite the COVID-19 pandemic we conducted face-to-face interviews while following the COVID-19 prevention guidelines. This was similar in the facilities and at the district where the primary method was observation. Through the field participant observation, the first author gathered first-hand information and particular attention was given to work environment (offices), work processes (procedures), data collection tools and report forms, use of support technology, data use, and supervision. Active engagement with the participants helped to get in-depth understanding of the data quality management practices, see **Figure 2.**

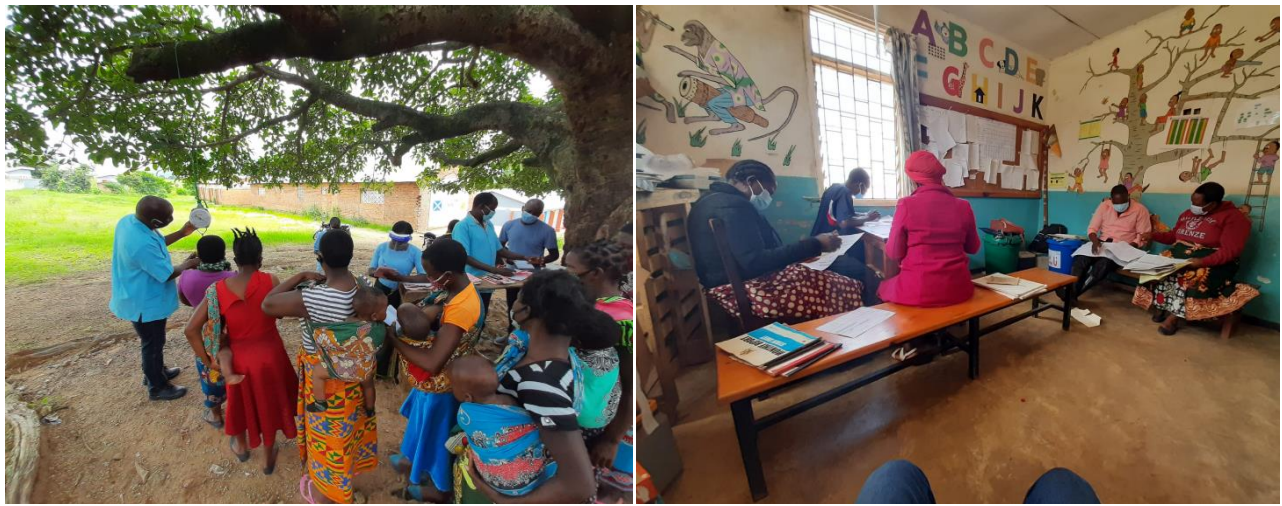

*Figure 2: Field work: Village Clinic and Report Production (Group)*

Observation and study of artefacts happened from the Village Clinics run by HSAs, Health Centres to Hospital at the district. The study of artefacts included performance reports from DHIIS2, health passport books, health registers, reporting forms both for capturing into DHIS2 and use at the facility. In the study of artefacts attention was on condition of HP book, actions being taken when there is a problem, care of the registers, records clarity and correctness. We did a random assessment of the previous reports submitted between December 2020 and January 2021 to validate the correctness and check the clarity of data. This helped to ascertain what the participants were saying against what they were actually doing.

### 3.3. DATA ANALYSIS

The study used qualitative data analysis applying both inductive and deductive strategy where codes were based on institutional theoretical constructs, legitimacy and isomorphism. We based on three types of legitimacy: pragmatic, moral and cognitive (Freitas & Guimarães, 2007; Bridwell-Mitchell & Mezias, 2012) . On isomorphism we used coercive, normative and mimetic isomorphism as our initiating codes (Currie, 2012; Kezar & Bernstein-Sierra, 2019). However, we also acknowledged the interesting emerging codes. We used a frequency table to help us determined important codes basing on frequency. However, pertinent codes were not just being left out regardless of their low frequency. Later we analyzed the relationship among the codes to build themes. These provided us with concepts of legitimacy as being experienced by the stakeholders in HMIS. After this we used the three isomorphic mechanisms to code isomorphic pressures being exerted in the HMIS. Again, by analyzing the relationship among the second group of codes, we established themes constituting the isomorphic pressure being experienced. Finally we analyzed the relationship between the legitimacy and isomorphic themes, see **Table 2.**

|  |  | Legitimacy themes | | |
|---|---|---|---|---|
|  |  | Moral | Pragmatic | Cognitive |
| **Isomorphism** | Normative |  |  | √ |
|  | Mimetic | √ | √ |  |
|  | Coercive |  |  |  |

*Table 2 Classification and relationship of themes*

## 4. FINDINGS

In regard to legitimacy we identified two sets of themes relating to data quality management practices. First meso level, which related to District Health Office and District Health Information System Office. Second, the micro level which related to medical facilities: Hospital, Health centers and Village Clinics which are manned by Health Surveillance Assistants. **Table 3** presents the legitimacy themes which were further classified into three: pragmatic, moral and cognitive.

| *Thematic Set* | **Legitimacy Type** | **Theme** |
|---|---|---|
| *1- Meso Level* | Pragmatic | i. Significance of data |
|  | Cognitive | ii. Implementation mechanisms |
| *2- Micro Level* | Pragmatic | iii. Individual conviction |
|  |  | iv. Shared responsibility |
|  | Moral/ Pragmatic | v. Significance of data |
|  | Cognitive | vi. Capitalizing existing structures |

*Table 3 List of legitimacy themes recognized in the findings*

We identified three isomorphic themes which were classified into two: mimetic and normative, see **Table 4.** We noted that these three themes relating to isomorphic forces played a role in helping legitimizing the data quality management practices. These are (a) policy and contractual obligations, (b) peer influence and (c) individual conviction.

| **Isomorphic Force Type** | **Theme** |
|---|---|
| **Normative** | Policy and contractual obligations |
| **Mimetic** | Peer influence |
|  | Individual conviction |

*Table 4 List of isomorphic themes recognized in the findings*

## 4.1. LEGITIMACY THEMES

The subsequent subsections provide a detailed explanation of the themes relating to legitimacy attributes stakeholders experienced the in data management practices.

i. **SIGNIFICANCE OF DATA**
Observations of the health service delivery process showed that participants value and use data in their work process. Notable observations include determining proper advice to mothers basing on their children's growth graph. Other instances include determining next schedule for women vaccine. *"You were not supposed to come today for vaccination, come on 18th March"*. These observations were made from two health centres and including two village clinics in each of the health centres. It involved under five and family planning clinics. *"Data is very important, it help us is our programming"*. This theme was prevalent at both the meso and micro level.

ii. **IMPLEMENTATION MECHANISMS**
The findings showed that stakeholders through government, Ministry of Health, have put in place structures to support data quality management practices across these key levels in data collection. Key are establishment of positions to support data management like Data Clerks both in Health Centres and Hospitals, and Health Management Information Systems Officers at district level.

iii. **INDIVIDUAL CONVICTION**

It was interesting to see personal effort among the participant especially among Data Clerks and Health Centre In-charges. *"I have never done training since I left school, but I just put effort to learn how to use these registers and produce reports". "It is painful that at times we have to use our own resources to buy internet data so that we can send reports. They promise to reimburse but two months ago we were not reimbursed up to now"* The same was noted among Health Surveillance Assistants, who had said that they rarely get supervised but always attempt to ensure proper record management arguing *"Its inhuman not to do the right thing just because you are not being supervised, that's spiritually immoral"*.

iv. **SHARED RESPONSIBILITY**
Unlike the rest of the themes, this theme was based on a code with less frequency. This, we noted, was due to the diverse approach used by the Health Centres in report generation. However, it was interesting because with the team approach, we recognised among the participants, the will to criticise and correct each other in attempt to ensure that they submit reliable reports. *"We try to avoid queries from the district and if queries come we take the blame on us all"*.

v. **CAPITALIZING EXISTING STRUCTURES**
This theme was drawn from two observations; the use of Health Surveillance Assistants (HSAs) as primary data collectors and positioning of data collection points (Data Clerks) within the medical process of diagnosis and prescription. For instance, Data Clerks were usually positioned between Clinicians and pharmacies. Similar observations were made in maternal health sections where In-charge of sections were made Focal Persons for data collection and reporting.

## 4.2. ISOMORPHIC THEMES

This subsection explains in detail themes relating to isomorphic pressures being exerted on stakeholders who are involved in data management practices. As illustrated in **Table 4** these we classified into mimetic and normative isomorphism.

i. **POLICY AND CONTRACTUAL OBLIGATIONS**

The theme was drawn from two observations, first to do with establishment of structures which are supporting the data management in medical facilities like Data Clerks and HMIS Officers which is seen in government policies. Second was among Focal Persons who work with Health Partners in various programmes. The finding showed that the Focal Persons were obliged to satisfying contract obligations they had with the Health Partners in respective programmes.

ii. **PEER INFLUENCE**

In one of our Village Clinics visit, we appreciated the role that Senior Health Surveillance Assistants played in guiding others into ensuring proper record management. *"Mr X, during our last meeting we discussed well these same shortfalls but it seems that here we are experiencing the same problems. When we assigned a role to oversee data collection in this catchment area, we recognised your potential and trusted that will help manage these problems"*. Similar observations were made in other areas and another notable one was at Centre Y where the reports were being generated during a meeting. Members would query each other and clarifications or corrections were being made.

iii. **INDIVIDUAL CONVICTION.**

From the HSAs, Data Clerks to Health Centre In-charge we noted a personal will to ensure data was collected, processed and sent in time. Despite the challenges which they complained like lack of training and support, they all in their own way made sure they produced and submitted the reports. Each of them individually or as a group had a workaround. *"I orient myself"*, *"we just train them on job"*, *"I sometime end up using my own airtime but not often"*, such were the common responses.

## 5. DISCUSSION AND CONCLUSION

The demand for quality data has been a long time aspiration in health service delivery in Malawi both at operational and strategic levels. Deliberate policies have been implemented over the past decade to support actualization of data quality management practices and subsequently realization of quality data (Ministry of Health -Malawi, 2018). Much as we appreciate the progress realized so far, this study contributes towards those efforts by explaining how the existing practices in the medical profession are contributing towards legitimation of the data quality management practices from health centers to districts.

### 5.1. LINKING ISOMORPHISM AND LEGITIMACY

Researchers have attempted to answer the question whether isomorphism leads to legitimacy (Freitas & Guimarães, 2007). The findings in this study are similar to prior studies and agree to those findings which suggested that that isomorphism play a role in legitimacy (DiMaggio & Powell, 1983). Although there is a continued debate on whether legitimacy leads to adherence, in this paper we will discuss our finding regarding the how isomorphic forces lead to legitimacy of data management practices in HMIS. **Figure 3** illustrates the link between these two concepts.

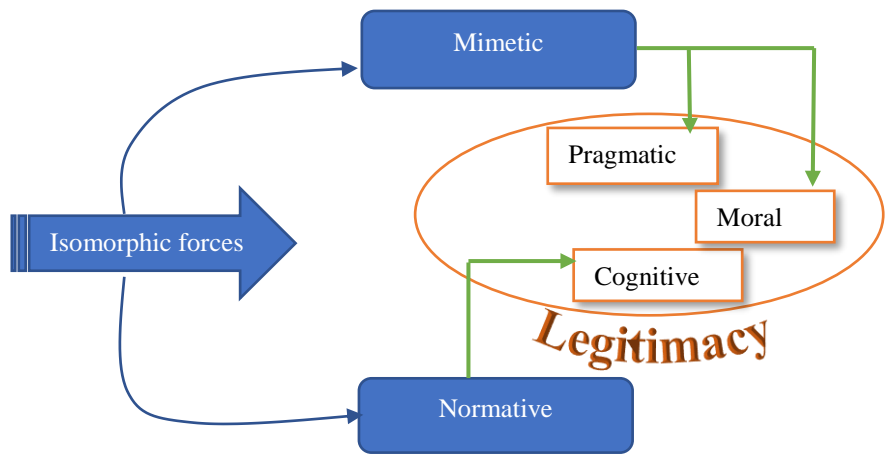

*Figure 3:* *Link between Isomorphic Forces and Legitimacy*

Data quality management practices guide those involved in managing the data; they are defined in policies and standard operating procedures. However, these practices are finding users who are already used to their own ways of working and having their own focus of attention. In case of health care delivery the focus is inevitably treatment of the patients. Bringing these data quality management practices may come in as intrusive in their working behavior. However, environmental conditions may help the players to accommodate the practices, such that over time the practices become socially accepted within the members' working culture. That's becoming legitimate (Randrianasolo & Randrianasolo, 2017). This legitimacy, from our findings trickles down from top to bottom, for example, the Ministry of Health created Data Clerk positions (establishments) as one of the organizational fields which were positioned within the health service delivery chain. We argue that by establishing the role of Data Clerk it signifies the acceptance of this role as an essential role within the ministry and within care delivery in particular. This we recognized as cognitive legitimacy realized through availability of implementation mechanisms, see **Table 3**. Further to this when these Data Clerks execute their duties, they are bound to work within their job descriptions as specified in the originating policies. We hence posit that a kind of normative isomorphism comes into play to influence the legitimization of this organizational field (Data Clerk) as these and practitioners like Directors of Health and Social Services (DHSS), formerly District Health Officers- (DHOs) are bound by policy to ensure the roles expectations are met. The same kind isomorphism is seen in legitimization of data quality management practices among Focal Persons who are bound to adopt these practices in an attempt to satisfy contractual obligations agreed with Health Partners in their various programmes, see **Table 4.**

The narration above would be similar to capitalization of existing structures as form of Cognitive legitimacy. A good example is where we see the use HSAs or Clinicians as Focal Persons for data collection. What we see is that by using existing roles it is made easy for the data management practices to be embedded within their working practices. With this in mind we notice some form of normative isomorphism since these are obliged to work within some predefined norms or standards.

We then discuss significance of data as a kind of moral legitimacy first among HSAs who deliver community health care services mostly in village clinics and end up developing relationships with the community members. To them they see it that they ought to have quality data so that they can deliver quality services to the members of the community. They have demonstrated that they use the data in their day to day activities and that they find it rather inhuman to deprive a fellow human a proper service just because they did not keep proper records. Hence as they work they do their best, regardless of challenges, to ensure they collect data and use the data. Although others may argue otherwise, we see some kind of mimetic isomorphism just because this is more or less personal conviction than based on some kind of standard or regulation. On the other hand, we see significance of data as a pragmatic legitimacy which is being driven by some form normative isomorphism. This time we relate it to the HMIS Officers and the DHSS who are pressurized through policy and operating standards to produce reports to support their planning and performance in their management of health care service delivery within their districts.

We finish our discussion of the link between legitimacy of data quality management practices and isomorphic forces with a reflection on shared responsibility as some kind of pragmatic legitimacy. This as presented in Section 4.1, is prevalent among HSAs, who appreciate the significance of team work. They perform data validation among themselves and they ensure each one produce reports by the said date. This is just their own workaround and is not enshrined in any policy governing their role. In this, we envisage a mimetic force emanating from peer pressure as individuals try to avoid letting down others. In the long run, we see that through team working data management practices become part of their work.

This understanding of the link between legitimacy and isomorphic pressure in HMIS using DHIS2, we argue can help to determine actions aimed at improving data quality. Our understanding is similar to the work of Rabearivony, Eloi, Jules, & Thorstrom (2008) who emphasized the importance of understanding informal institutional environment arguing that it is crucial in reaching to objectives, which in our case is quality data for improving health service delivery.

## 5.2. DATA QUALITY AND LEGITIMACY

We have argued in Section 5.1 that isomorphic forces play a role in legitimization of data quality management practices. Further, in Section 1 we defined data quality as how data satisfy expectations of the users, we further pointed out that among the dimensions of data quality, the study opted to focus on correctness and timeliness. Basing on this understanding we now explain how legitimization of data quality management practices by institutions may lead to data quality as illustrated in Figure 4.

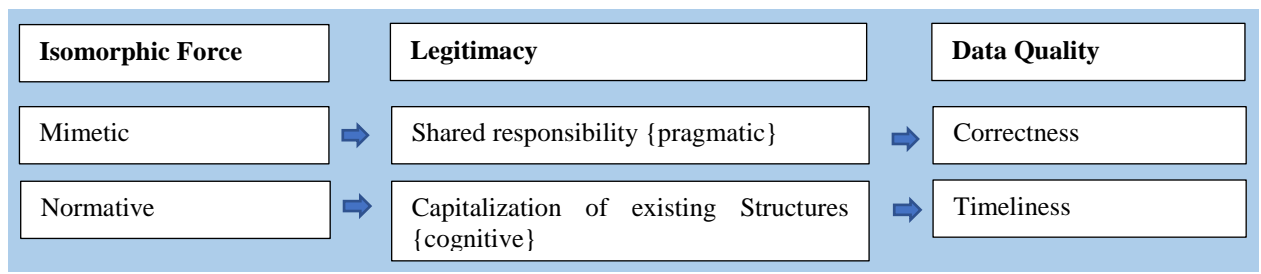

*Figure 4: Linking Legitimacy and Data Quality*

Firstly, we look at how the legitimization leads to correct data (correctness) among HSAs. We state that through shared responsibility which we related to pragmatic legitimacy, the HSAs correct or validate their data especially when they are producing reports. The reports in turn help to satisfy their expectations, for example they highlighted getting the right vaccine and family planning doses. The HSAs indicated that they have to give correct information because it affects the supplies which are based on their reports. During the study, we noted the HSAs strived to make sure they were as correct as possible specific data. For example, these are indicators for vaccine: quantity received, quantity administered, and quantity damaged.

Secondly, we look at capitalization of existing structures which we related to cognitive legitimacy. In this we see the Health Partner engaging nurses or clinicians as Focal Persons in reporting. In turn, the Focal Persons are forced to satisfy the contractual obligations and in the long run the Health Partners get their reports in time for timely decision making. These in essence shows that legitimacy may help to meet some of the characteristic of data quality like correctness and timeliness.

## 6. CONCLUSSION

Existing health professional practices and structures provide two isomorphic forces (mimetic and normative) that are contributing to legitimization of data quality management practices in health management information systems. Subsequently, the legitimization leads to achieving data quality especially correctness and timeliness. However, study findings did not establish presence of coercive isomorphism. This we attributed to the sidelining of key regulatory bodies in the medical field: the Nurses and Midwives Council of Malawi (NMCM) and the medical Council of Malawi (MCM). We hence argue for further studies to explore how these influential bodies can play a role in influencing legitimization and adoption of the data quality management practices. Their presence we posit can help to go beyond mere legitimacy to accountability of key players in data quality management within HMIS hence increasing data quality in routine health information especially those using DHIS2 platform.

This study is part of a longitudinal study and the field work only involved one of the two districts ear marked for the entire study. As such, the findings may lack a comparative view of the experiences in the other areas. However, we feel that the findings of the study provide a foundation to explore further the issues of legitimacy and isomorphism and indeed their application to data quality management practice in HMIS. Another concern was that some of participants who participated in the study were newly recruited and may have less experience to share, nonetheless, we are content that the wider pool of participants helped to mitigate the shortfalls from this concern.

# REFERENCES AND CITATIONS